\documentclass{article}

    \usepackage[final]{neurips_2020}

\usepackage[utf8]{inputenc} 
\usepackage[T1]{fontenc}    
\usepackage{hyperref}       
\usepackage{url}            
\usepackage{booktabs}       
\usepackage{amsfonts}       
\usepackage{nicefrac}       
\usepackage{microtype}      
\usepackage[pdftex]{graphicx}

\title{White-box Audio VST Effect Programming}

\author{%
  Christopher Mitcheltree \\
  Tokyo Institute of Technology \\
  Qosmo Inc. \\
  \texttt{christhetree@gmail.com} \\
   \And
   Hideki Koike \\
   Tokyo Institute of Technology \\
   \texttt{koike@c.titech.ac.jp} \\
}

\begin{document}

\maketitle

\begin{abstract}
  Learning to program an audio production VST plugin is a time consuming process, usually obtained through inefficient trial and error and only mastered after extensive user experience. We propose a white-box, iterative system that provides step-by-step instructions for applying audio effects to change a user's audio signal towards a desired sound. We apply our system to Xfer Records Serum: currently one of the most popular and complex VST synthesizers used by the audio production community. Our results indicate that our system is consistently able to provide useful feedback for a variety of different audio effects and synthesizer presets.
\end{abstract}

\begin{figure}[h]
  \centering
  \includegraphics[width=1.0\linewidth]{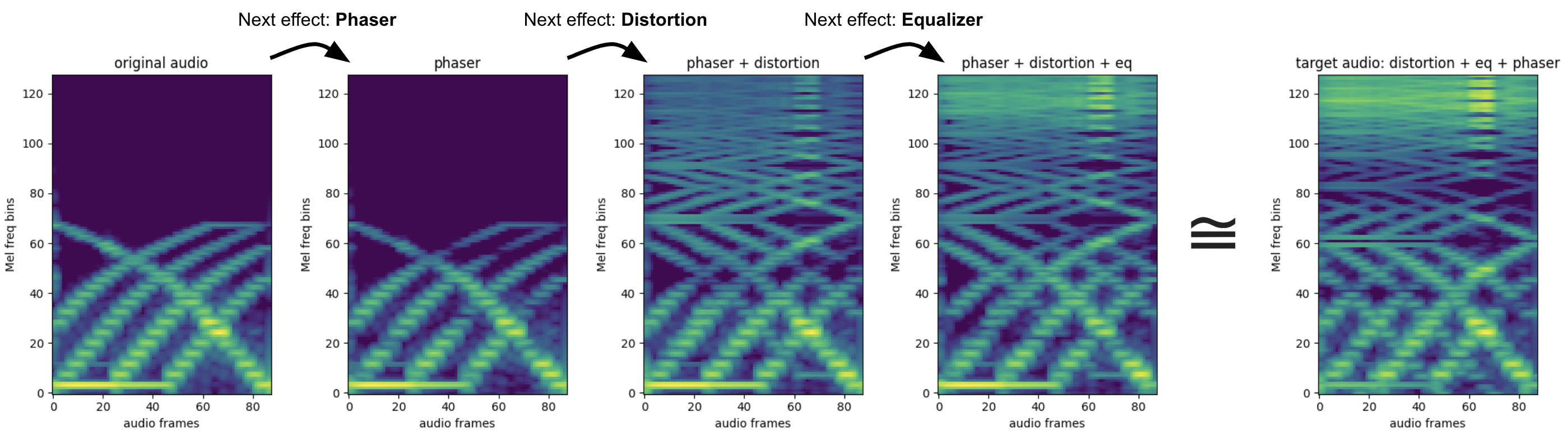}
  \caption{Mel spectrogram progression of our system applying three effects to a user's input audio.}
\end{figure}

\section{Introduction and Background}

Sound design is the process of using a synthesizer and audio effects to craft a desired output sound, typically by leveraging virtual studio technology (VST) on a computer. Often, the audio effects applied to the synthesizer play the biggest role in producing a desired sound. Sound design for the music industry is a very difficult task done by professionals with years of experience. Educational tools are limited and beginners are usually forced to learn via trial and error or from online resources created by others more experienced who typically also learned in a similar way.

Prior work leveraging AI to program audio VSTs uses genetic algorithms [1; 2; 3], genetic programming [3], k-means clustering + tree-search [4], and deep convolutional neural networks [2; 5] to achieve this objective. There has also been research on using deep learning to model or apply audio effects directly [6; 7; 8; 9]. These systems typically suffer from one or more of the following problems: they are applied to toy VSTs with little practical use, they are incompatible with existing VSTs, their inference time is prohibitively long, or they are black-boxes with uninterpretable results.

When using an AI assisted system, the user's sense of ownership over their work should be preserved. Our system is inspired by white-box automatic image post-processing systems [10] and collaborative production tools [11; 12] that can educate and augment a user rather than aiming to replace them. 

\section{System Overview}

Our system iteratively nudges an input audio towards the same timbre of a desired target audio and provides interpretable intermediate steps. It uses, to the best of our knowledge, a novel approach consisting of an ensemble of models working together: a recurrent neural network (RNN) to select which effect to apply next and then a collection of convolutional neural networks (CNN), one per supported effect, to apply the correct effect parameters. An example sequence of spectrograms and steps output by our system is shown in Figure 1.

We collect training data using five timbre-changing audio effects from Serum's effects rack: multi-band compression, distortion, equalizer (EQ), phaser, and hall reverb. We also use 12 different synthesizer presets split into three groups: \emph{basic shapes} (sine, triangle, saw, and square wave), \emph{advanced shapes} (four presets), and \emph{advanced modulating shapes} (four presets).

Since our system focuses on reproducing a desired timbre, we represent input audio as power dB Mel spectrograms. Using a modified automated VST rendering tool [13], \textasciitilde120k one second long audio clips are generated for each synthesizer preset and are sampled from all possible combinations of the five supported effects.

The CNN effect models take as input two spectrograms stacked together (two channels total): the target spectrogram and the input spectrogram. Their outputs vary depending on the effect they are modeling, but consist of some combination of binary, categorical, and continuous outputs. A Cartesian product is used for selecting current and target spectrograms to train on, thus resulting in \textasciitilde1.2M available training data points for each effect.

The RNN model takes as input an arbitrary length sequence consisting of the CNN effect model spectrogram input and a sequence of one hot vectors of the same length representing the used effects. Its output is a 5-dimensional softmax layer indicating the probability of the next effect to be applied. 

More details about data collection, model architectures, and training can be found in supplemental Figures 2 and 3 and supplemental Table 2. 

\section{Evaluation and Discussion}

\begin{table}[h]
  \caption{Mean errors and $\Delta$s for input and output audio from the \emph{basic shapes} preset group.}
  \centering
  \begin{tabular}{l}
    \toprule
    \multicolumn{1}{c}{} \\
    \cmidrule(r){1-1}
    Metric \\
    \midrule
    MSE \\
    MAE \\
    MFCC \\
    LSD \\
    \bottomrule
  \end{tabular}
  \begin{tabular}{lll}
    \toprule
    \multicolumn{3}{c}{Mean Error against Target Audio} \\
    \cmidrule(r){1-3}
    Init. Audio & Final Audio & $\Delta$ \\
    \midrule
    0.055 & 0.012 & -0.043 \\
    0.172 & 0.074 & -0.098 \\
    157.15 & 70.06 & -87.09 \\
    16.16 & 7.62 & -8.54 \\
    \bottomrule
  \end{tabular}
  \begin{tabular}{lllll}
    \toprule
    \multicolumn{5}{c}{Mean Error $\Delta$ per Step} \\
    \cmidrule(r){1-5}
    1 & 2 & 3 & 4 & 5 \\
    \midrule
    -0.024 & -0.013 & -0.008 & -0.004 & -0.002 \\
    -0.052 & -0.034 & -0.020 & -0.008 & -0.004 \\
    -45.97 & -28.64 & -20.13 & -7.30 & -4.28 \\
    -4.62 & -3.04 & -1.64 & -0.60 & -0.32 \\
    \bottomrule
  \end{tabular}
\end{table}

The CNN effect models are evaluated individually against their parameter reconstruction ability and how closely their output matches the target audio. Audio similarity is measured via four different metrics: MAE and MSE between the two power dB spectrograms, and the mean Euclidean distance between the first 20 Mel frequency cepstral coefficients (MFCC) and the log-spectral distance (LSD) between the two power spectrograms. The RNN model is evaluated against its prediction accuracy for the next effect and the entire ensemble of models is evaluated against changes in audio similarity as steps are taken by the system. Evaluation results for our entire system are shown in Table 1 and additional results can be found in supplemental Tables 3, 4, 5, and 6.

The results indicate that our system is consistently able to produce intermediate steps that bring the input audio significantly closer to the target audio.\footnote{Audio examples can be listened to at \url{https://bit.ly/serum_rnn}} Our system also provides near real-time, quantitative feedback about which effects are the most important. The user can pick and choose which intermediate steps they would like to use and can feed tweaked versions back into the system for additional fine-tuning or to learn more. We also noticed fun creative applications when our system produced unexpected results or was given significantly out of domain target audio. 

\section{Ethical Implications}

Combining artificial intelligence with creativity carries with it various different ethical considerations, one of which is a potential future decrease in demand for professional audio producers due to an increasing ability to replace them with technology. We believe the best approach to this is to build systems that are collaborative and can augment people rather than replacing them entirely. While we believe our research is just the tip of the iceberg for building AI powered sound design tools, we can imagine a future where tools like ours might be able to find more efficient and simpler methods of creating sounds, thus educating students more effectively and democratizing sound design. We compare this to a similar situation that occurred when beat-matching technology was invented and added to DJ systems (to the disgruntlement of some DJing "purists"). However, this sometimes controversial technology democratized DJing and enabled a new generation of artists to focus on new creative applications, thus progressing the community as a whole.

\section{Acknowledgements}

We would like to thank Erwin Wu for providing additional computing resources.

\section{References}
\small

[1] Tatar, K., Matthieu Macret and P. Pasquier. “Automatic Synthesizer Preset Generation with PresetGen.” \emph{Journal of New Music Research 45} (2016).

[2] Yee-King, Matthew, Leon Fedden and Mark d'Inverno. “Automatic Programming of VST Sound Synthesizers Using Deep Networks and Other Techniques.” \emph{IEEE Transactions on Emerging Topics in Computational Intelligence 2} (2018).

[3] Macret, Matthieu and P. Pasquier. “Automatic design of sound synthesizers as pure data patches using coevolutionary mixed-typed cartesian genetic programming.” \emph{GECCO '14} (2014).

[4] Cáceres, Juan-Pablo. “Sound Design Learning for Frequency Modulation Synthesis Parameters.” (2007).

[5] Barkan, Oren, David Tsiris, O. Katz and Noam Koenigstein. “InverSynth: Deep Estimation of Synthesizer Parameter Configurations From Audio Signals.” \emph{IEEE/ACM Transactions on Audio, Speech, and Language Processing 27} (2019).

[6] Ramírez, M. A. M. and J. Reiss. “End-to-end equalization with convolutional neural networks.” \emph{International Conference on Digital Audio Effects} (2018).

[7] Damskägg, Eero-Pekka, Lauri Juvela, Etienne Thuillier and V. Välimäki. “Deep Learning for Tube Amplifier Emulation.” \emph{ICASSP 2019 - 2019 IEEE International Conference on Acoustics, Speech and Signal Processing (ICASSP)} (2019).

[8] Sheng, Di and Gyorgy Fazekas. “A Feature Learning Siamese Model for Intelligent Control of the Dynamic Range Compressor.” \emph{2019 International Joint Conference on Neural Networks (IJCNN)} (2019).

[9] Engel, J., Lamtharn Hantrakul, Chenjie Gu and Adam Roberts. “DDSP: Differentiable Digital Signal Processing.” \emph{2020 International Conference on Learning Representations (ICLR)} (2020).

[10] Hu, Yuanming \& He, Hao \& Xu, Chenxi \& Wang, Baoyuan \& Lin, Stephen.  "Exposure: A White-Box Photo Post-Processing Framework." \emph{ACM Transactions on Graphics.} (2017).

[11] Sommer, Nathan and A. Ralescu. “Developing a Machine Learning Approach to Controlling Musical Synthesizer Parameters in Real-Time Live Performance.” \emph{MAICS} (2014).

[12] Thio, Vibert, and Chris Donahue. "Neural Loops." \emph{2019 NeurIPS Workshop on Machine Learning for Creativity and Design} (2019).   

[13] Fedden, Leon. "RenderMan". GitHub. \url{https://github.com/fedden/RenderMan} (accessed 2020).

\pagebreak
\normalsize

\section{Supplemental Material}

\subsection{Data Collection}

\begin{table}[h]
  \caption{Parameters sampled from the Serum VST synthesizer.}
  \centering
  \begin{tabular}{llll}
    \toprule
    Effect & Parameter Name & Type & Sampled Values \\
    \midrule
    Compressor & Low-band Compression & Continuous & [0.0, 1.0] \\
    Compressor & Mid-band Compression & Continuous & [0.0, 1.0] \\
    Compressor & High-band Compression & Continuous & [0.0, 1.0] \\
    Distortion & Mode & Categorical & 12 classes \\
    Distortion & Drive & Continuous & [0.3, 1.0] \\
    Equalizer & High Frequency Cutoff & Continuous & [0.50, 0.95] \\
    Equalizer & High Frequency Resonance & Continuous & [0.0, 1.0] \\
    Equalizer & High Frequency Gain & Continuous & [0.0, 0.4] and [0.6, 1.0] \\
    Phaser & LFO Depth & Continuous & [0.0, 1.0] \\
    Phaser & Frequency & Continuous & [0.0, 1.0] \\
    Phaser & Feedback & Continuous & [0.0, 1.0] \\
    Hall Reverb & Mix & Continuous &  [0.3, 0.7] \\
    Hall Reverb & Low Frequency Cutoff & Continuous &  [0.0, 1.0] \\
    Hall Reverb & High Frequency Cutoff & Continuous &  [0.0, 1.0] \\
    \bottomrule
  \end{tabular}
\end{table}

Data collection and processing systems represent a significant portion of the software engineering effort required for this project. Table 2 summarizes which Serum synthesizer parameters are sampled for each supported effect. Parameter sampling value ranges are occasionally limited to lie within practical, everyday use regions.

The \emph{basic shapes} preset group consists of the single oscillator sine, triangle, saw, and square wave default Serum presets.

The \emph{advanced shapes} preset group consists of the dry (no effects) dual oscillator \texttt{"LD Power 5ths"}, \texttt{"SY Mtron Saw"}, \texttt{"SY Shot Dirt Stab"}, and \texttt{"SY Vintage Bells"} default Serum presets.

The \emph{advanced modulating shapes} preset group consists of the dry dual oscillator \texttt{"LD Iheardulike5ths"}, \texttt{"LD Postmodern Talking"}, \texttt{"SQ Busy Lines"}, and \texttt{"SY Runtheharm"} default Serum presets. All of these presets also use intense time varying modulations.

Audio samples are played and rendered for one second using a MIDI pitch of C4, maximum velocity, and a sampling rate of 44100 Hz. In the future we would like to include audio pitch, attack, decay, sustain, and release features directly into our system by modifying these values.

Mel spectrograms are calculated using a hop length of 512 samples, a FFT window length of 4096, and 128 Mel filters.

\subsection{Modeling}

\begin{figure}[ht]
  \centering
  \includegraphics[width=0.85\linewidth]{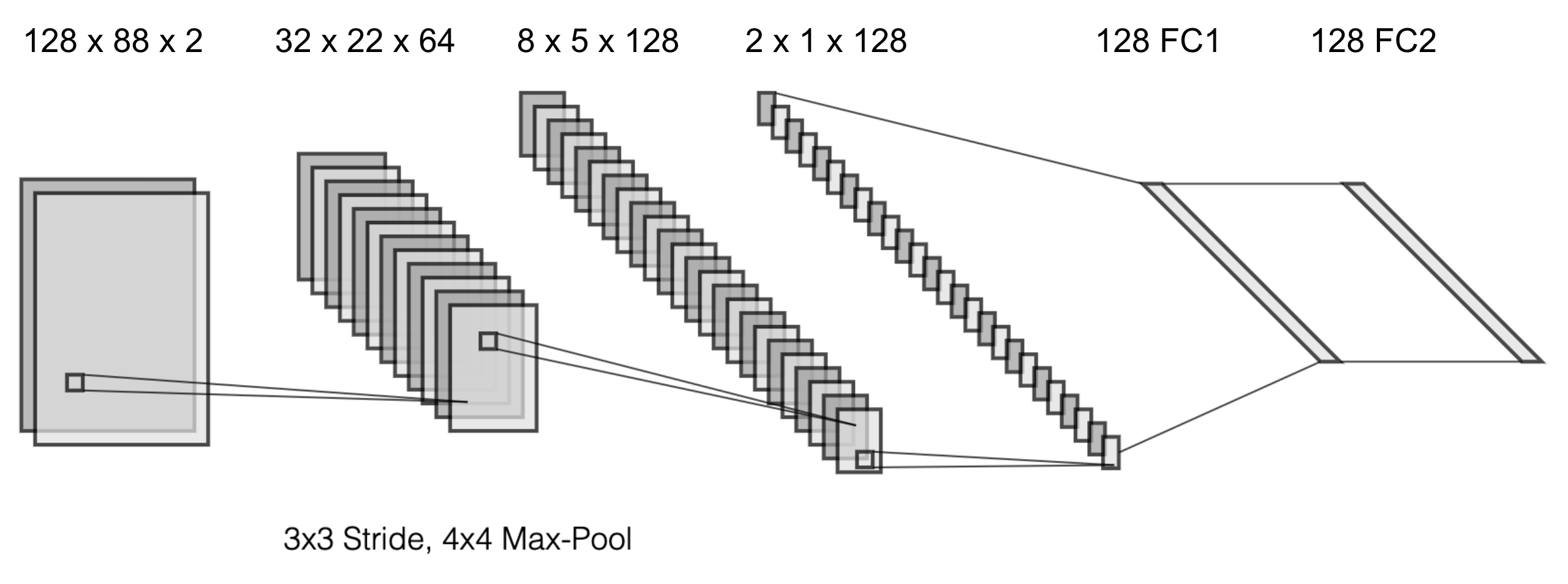}
  \caption{CNN effect model architecture (not to scale).}
\end{figure}

All five CNN effect models use ELU activations and a 50\% dropout rate for each of their fully connected (FC) layers. Their architecture is shown in Figure 2. They are trained using a batch size of 128, mean squared error loss for continuous parameters, binary cross-entropy loss for binary parameters, and categorical cross-entropy loss for categorical parameters. 

\begin{figure}[ht]
  \centering
  \includegraphics[width=0.72\linewidth]{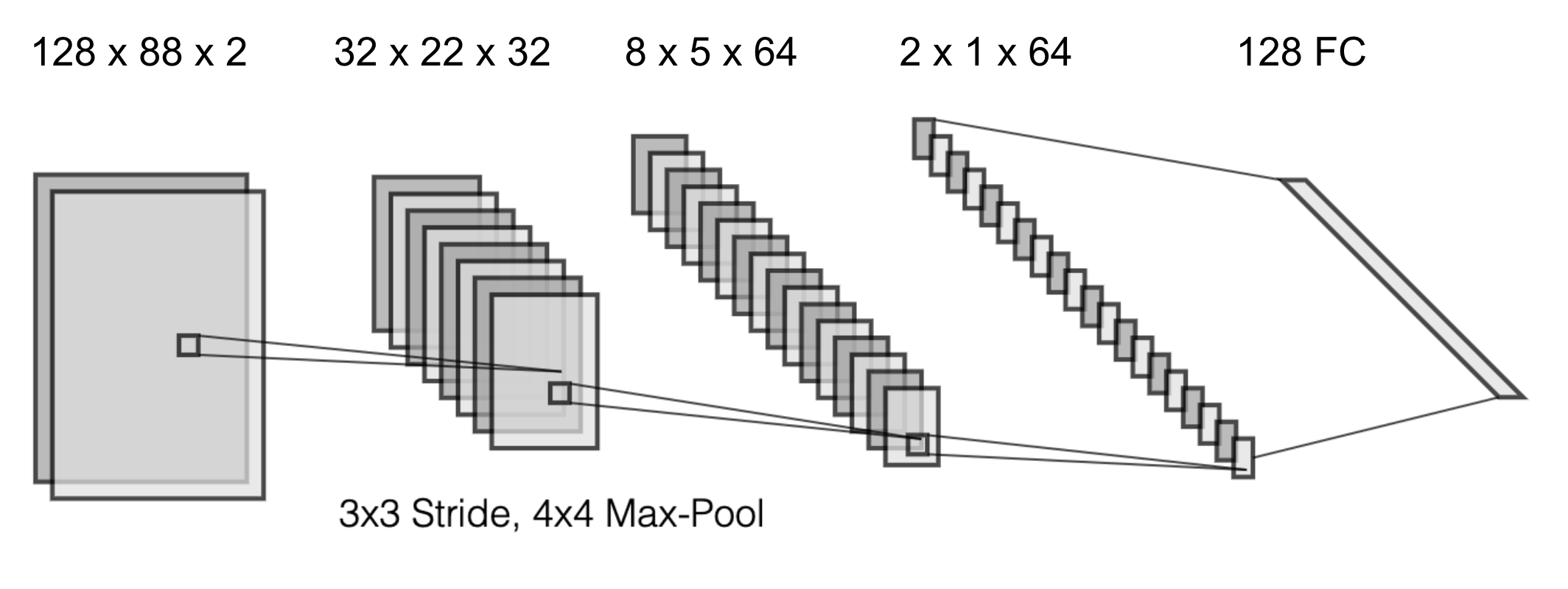}
  \caption{CNN model architecture used in the RNN next effect prediction model (not to scale).}
\end{figure}

The RNN model consists of a bi-directional, 128-dimensional LSTM layer followed by a 128-dimensional FC layer and lastly the 5-dimensional softmax output layer. The FC layer uses ELU activation units and a 50\% dropout rate. Features are extracted from the Mel spectrogram sequence input using a smaller, time-distributed CNN with an architecture displayed in Figure 3. This sequence of extracted, 128-dimensional Mel spectrogram features is concatenated with the sequence of one hot vectors representing which effects have been used and is then fed as input to the LSTM layer. The RNN model is trained with a batch size of 32.

All models are trained for 100 epochs with early stopping and a validation and test split of 0.10 and 0.05 respectively. The Adam optimizer is used with a learning rate of 0.001.

\subsection{Evaluation}

\begin{table}[h]
  \caption{CNN effect models mean error reduction for all three preset groups.}
  \centering
  \begin{tabular}{lllll}
    \toprule
    && \multicolumn{3}{c}{Mean Error $\Delta$ against Target Audio} \\
    \cmidrule(r){3-5}
    Effect & Metric & \emph{Basic Shapes} & \emph{Adv. Shapes} & \emph{Adv. Mod. Shapes} \\
    \midrule
    Compressor & MSE  & -0.012 & -0.013 & -0.007 \\
    Compressor & MAE  & -0.050 & -0.049 & -0.030 \\
    Compressor & MFCC & -53.19 & -54.66 & -37.42 \\
    Compressor & LSD  & -4.20  & -4.15  & -2.42  \\
    \midrule
    Distortion & MSE  & -0.036 & -0.019 & -0.037 \\
    Distortion & MAE  & -0.062 & -0.056 & -0.082 \\
    Distortion & MFCC & -60.50 & -60.70 & -88.40 \\
    Distortion & LSD  & -5.30  & -5.11  & -7.18  \\
    \midrule
    Equalizer & MSE   & -0.004 & -0.009 & -0.005 \\
    Equalizer & MAE   & -0.018 & -0.038 & -0.019 \\
    Equalizer & MFCC  & -24.63 & -45.20 & -29.93 \\
    Equalizer & LSD   & -1.31  & -3.27  & -1.54  \\
    \midrule
    Phaser* & MSE      & 0.002  & 0.000  & 0.002  \\
    Phaser* & MAE      & 0.005  & -0.002 & 0.008  \\
    Phaser* & MFCC     & 1.23   & -5.98  & 1.96   \\
    Phaser* & LSD      & 0.64   & -0.11  & 0.99   \\
    \midrule
    Hall Reverb & MSE  & -0.016 & -0.005 & -0.007 \\
    Hall Reverb & MAE  & -0.064 & -0.029 & -0.033 \\
    Hall Reverb & MFCC & -47.16 & -26.61 & -31.59 \\
    Hall Reverb & LSD  & -6.27  & -2.46  & -3.10  \\
    \bottomrule
  \end{tabular}
\end{table}

* It's important to note that Mel spectrograms do not represent phase information well. As a result, the error metrics used are less representative of the phaser effect's error. A positive error $\Delta$ may occur even when the predicted audio sample clearly sounds much closer to the target audio sample when compared to the initial audio sample. We plan to include phase information in future iterations of our system.

\bigskip
\bigskip

\begin{table}[h]
  \caption{RNN model next effect prediction accuracy for all three preset groups.}
  \centering
  \begin{tabular}{llll}
    \toprule
    & \multicolumn{3}{c}{Mean Next Effect Prediction Accuracy} \\
    \cmidrule(r){2-4}
    Step & \emph{Basic Shapes} & \emph{Adv. Shapes} & \emph{Adv. Mod. Shapes} \\
    \midrule
    1 & 0.997 & 0.993 & 0.997 \\
    2 & 0.983 & 0.985 & 0.989 \\
    3 & 0.981 & 0.979 & 0.983 \\
    4 & 0.973 & 0.988 & 0.977 \\
    5 & 0.999 & 1.000 & 0.997 \\
    \midrule
    All & 0.983 & 0.985 & 0.986 \\
    \bottomrule
  \end{tabular}
\end{table}

\bigskip
\bigskip

\begin{table}[h]
  \caption{Mean errors and $\Delta$s for input and output audio from the \emph{advanced shapes} preset group.}
  \centering
  \begin{tabular}{l}
    \toprule
    \multicolumn{1}{c}{} \\
    \cmidrule(r){1-1}
    Metric \\
    \midrule
    MSE \\
    MAE \\
    MFCC \\
    LSD \\
    \bottomrule
  \end{tabular}
  \begin{tabular}{lll}
    \toprule
    \multicolumn{3}{c}{Mean Error against Target Audio} \\
    \cmidrule(r){1-3}
    Init. Audio & Final Audio & $\Delta$ \\
    \midrule
    0.039 & 0.009 & -0.030 \\
    0.150 & 0.067 & -0.083 \\
    146.79 & 62.25 & -84.54 \\
    14.13 & 6.71 & -7.42 \\
    \bottomrule
  \end{tabular}
  \begin{tabular}{lllll}
    \toprule
    \multicolumn{5}{c}{Mean Error $\Delta$ per Step} \\
    \cmidrule(r){1-5}
    1 & 2 & 3 & 4 & 5 \\
    \midrule
    -0.017 & -0.010 & -0.007 & -0.003 & 0.000 \\
    -0.042 & -0.030 & -0.022 & -0.009 & -0.001 \\
    -43.45 & -30.36 & -21.87 & -9.00 & -0.40 \\
    -3.91 & -2.66 & -1.83 & -0.70 & -0.05 \\
    \bottomrule
  \end{tabular}
\end{table}

\bigskip
\bigskip

\begin{table}[h]
  \caption{Mean errors and $\Delta$s for input and output audio from the \emph{advanced mod. shapes} preset group.}
  \centering
  \begin{tabular}{l}
    \toprule
    \multicolumn{1}{c}{} \\
    \cmidrule(r){1-1}
    Metric \\
    \midrule
    MSE \\
    MAE \\
    MFCC \\
    LSD \\
    \bottomrule
  \end{tabular}
  \begin{tabular}{lll}
    \toprule
    \multicolumn{3}{c}{Mean Error against Target Audio} \\
    \cmidrule(r){1-3}
    Init. Audio & Final Audio & $\Delta$ \\
    \midrule
    0.049 & 0.013 & -0.036 \\
    0.181 & 0.077 & -0.104 \\
    176.37 & 72.52 & -103.85 \\
    16.90 & 7.74 & -9.16 \\
    \bottomrule
  \end{tabular}
  \begin{tabular}{lllll}
    \toprule
    \multicolumn{5}{c}{Mean Error $\Delta$ per Step} \\
    \cmidrule(r){1-5}
    1 & 2 & 3 & 4 & 5 \\
    \midrule
    -0.022 & -0.010 & -0.008 & -0.000 & -0.002 \\
    -0.070 & -0.026 & -0.018 & -0.004 & -0.004 \\
    -68.97 & -24.92 & -18.50 & -4.26 & -3.49 \\
    -6.17 & -2.22 & -1.49 & -0.33 & -0.32 \\
    \bottomrule
  \end{tabular}
\end{table}

\end{document}